\documentclass[hyper,letterpaper,12pt]{article}
\usepackage{a4wide}
\usepackage{amsmath}
\usepackage[
      colorlinks=true,
      linkcolor=blue,
      urlcolor=blue,
      filecolor=black,
      citecolor=blue,
            ]{hyperref}
\usepackage{color}
\usepackage{graphicx}
\usepackage{amsfonts}
\usepackage{amssymb}
\usepackage{epsfig}
\usepackage{subfigure}
\usepackage{amsmath}
\usepackage{latexsym}
\usepackage{mathrsfs}
\newcommand{\be}{\begin{equation}}
\newcommand{\ee}{\end{equation}}
 \newcommand{\bea}{\begin{eqnarray}}
 \newcommand{\ena}{\end{eqnarray}}

\begin{document}
\title{\bf \Large Weak cosmic censorship conjecture with pressure and volume in the Gauss-Bonnet AdS black hole}

\author{\large
~~Xiao-Xiong Zeng$^{1,2}$\footnote{E-mail: xxzengphysica@163.com},
~~Xin-Yun Hu$^{3}$\footnote{E-mail: huxinyun@126.com},
~~Ke-Jian He $^4$\footnote{E-mail: kjhe94@163.com}
\date{\today}
\\
\\
\small $^1$ Department of Mechanics, Chongqing Jiaotong University, Chongqing ~400074, China\\
\small $^2$  State Key Laboratory of Mountain Bridge and Tunnel Engineering, Chongqing Jiaotong \\
\small University, Chongqing 400074, China \\
\small $^3$ College of Economic and Management, Chongqing Jiaotong University,\\
\small Chongqing 400074, China\\
\small $^4$ Physics and Space College, China West Normal University, Nanchong 637000, China}

\maketitle

\begin{abstract}
\normalsize With the  Hamilton-Jacobi equation, we obtain the energy-momentum relation of a charged particle as it is absorbed by the Gauss-Bonnet AdS black hole. On the basis of the energy-momentum relation at the event horizon,
we investigate the first law, second law, and weak cosmic censorship conjecture
in both the normal phase space and extended phase space. Our results show that the first law, second law as well as the weak cosmic censorship conjecture are  valid in the normal phase space for all the initial states  are black holes. However, in the extended phase space, the second law is violated for the extremal and near-extremal black holes, and the weak cosmic censorship conjecture is violable for the near-extremal black hole, though the first law is  still valid.  In addition, in both  the the normal and extended   phase spaces, we find   the absorbed particle changes the configuration of the near-extremal black hole,  while don't change that of   the extremal  black hole.
\end{abstract}
\newpage

%\tableofcontents

\section{Introduction}\label{sec1}
According to the singularity theorems developed by Penrose and Hawking \cite{Hawking:1969sw}, we know that the formation of a singularity with infinite matter density is inevitable
during the gravitational collapse. The existence of a singularity will destroy the deterministic nature of general relativity.
To circumvent this problem, Penrose thus claimed that the singularity produced in
the gravitational collapse must be hidden within a
black hole so that a distant observer cannot perceive it \cite{Penrose}, which is the so-called  weak cosmic censorship conjecture.   In this case, the  expected predictability and
deterministic nature of general relativity is reassured.

There is not  a concrete proof of the weak cosmic censorship
conjecture, we thus should check its validity in different spacetimes.  Wald proposed firstly a gedanken experiment to  check this conjecture
 by examining whether the black hole horizon
can be destroyed by injecting a point particle \cite{Wald:1974ge}.  For an extremal Kerr-Newman black hole, he found that   a particle which violate the  weak cosmic censorship
conjecture will not be absorbed by
the black hole.  Until now, there are some debates on the test particle model.  In the near-extremal   Reissner-Nordst\"{o}m black hole and Kerr black hole, the  cosmic censorship
conjecture was found to be violated  in \cite{Hubeny:1998ga} and \cite{Jacobson:2009kt} respectively.
As
 the higher order terms in the energy,
angular momentum, and charge of the test particle are taken into account, the weak  cosmic censorship
conjecture was found to be violated too   even for  an extremal Kerr-Newman black hole \cite{Gao:2012ca}.
Later, it was claimed that  in all
of these situations, the test particle assumption  was not perfect  since they did not  take into account the self force and back reaction effects \cite{Barausse:2010ka}. As these effects were considered, the weak cosmic censorship
conjecture was found to be valid for both the extremal  and near-extremal black holes \cite{Colleoni:2015afa}. Especially, by  applying the  Wald formalism rather than  matter, a new version of gedanken experiment has been designed recently \cite{Wald2018,Sorce:2017dst}.  The weak cosmic censorship
conjecture  was  found to be  valid for  the non-extremal black holes  \cite{Wald2018,Sorce:2017dst,Ge:2017vun,An:2017phb}.  In this framework,  the second order variation of the
mass of the black hole emerges, which somehow incorporates both the self force and back reaction effects.

With different methods, there  have been some counter examples to the weak cosmic censorship conjecture, especially in spacetimes
with more than four dimensions \cite{Santos:2015iua,Lehner:2010pn,Gregory:1993vy,Gim:2018axz}. Even in the four dimensional AdS  black holes, there are also some counter examples recently.   In the
Einstein-Maxwell theory \cite{Crisford2017}, the weak cosmic censorship conjecture was found to be violated since the curvature grows without bound in the future to the infinite boundary observers.  In the Einstein-Maxwell-dilaton theory, it was found that there was not a horizon covering the singularity and the singularity are
connected by the traversable wormholes \cite{Goulart:2018ckh}. The weak  cosmic censorship was found to be related with the weak gravity. It was suggested  that the above counterexamples  might be removed if the weak gravity
conjecture  holds \cite{Horowitz:2016ezu,Horowitz:2019eum,Yu:2018eqq,Crisford:2017gsb}.

Besides the weak cosmic censorship conjecture, the test particle model also can be used to discuss the first law and second law of thermodynamics of black holes. With the relation between the energy and momentum of the test particle, the first law and second law of a three  dimensional  black hole have been investigated \cite{Gwak:2015sua}.  The merit of the test particle model is that it also can be used to study the thermodynamics and weak cosmic censorship conjecture in the extended phase space, where
the cosmological parameter and its conjugate quantity are regarded as the pressure and volume respectively \cite{Dolan:2010ha,Cvetic:2010jb}.  In the extended phase space, the first law of thermodynamics and Van der Waals-like phase transition have been investigated extensively \cite{Dolan:2010ha,Cvetic:2010jb,Kastor:2009wy,Zeng:2015wtt,Zeng:2017zlm,Zeng:2016fsb,Mo:2016cmi,Zeng:2016aly}. However, there is little work to discuss the second law as well as the weak cosmic censorship conjecture. The validity of the first law dose not imply that the second law and  the   weak cosmic censorship conjecture are valid. Therefore,  it  is of great importance and necessity to study  the second law  and the   weak cosmic censorship in the extended phase space.
The laws of thermodynamics and  weak cosmic censorship conjecture with pressure and volume  in the high dimensional  Reissner Nordstr\"{o}m-AdS black hole have been investigated recently \cite{Gwak:2017kkt}. It was found that the first law and the weak cosmic censorship conjecture
were valid,
 while the second law   was violated  for the extremal and  near-extremal black holes, which is different from the case without pressure and volume.  Now, the idea in \cite{Gwak:2017kkt} has been extended to the Born-Infeld-anti-de Sitter black hole \cite{Zeng:2019jta,Wang:2019dzl},  torus-like AdS black hole \cite{Han:2019kjr},  and three dimensional BTZ black holes \cite{Zeng:2019jrh,Han:2019lfs}.  Especially, thermodynamics and weak cosmic censorship conjecture in the Kerr-AdS black hole have also been investigated \cite{Zeng:2019aao}.  Different from the spherically symmetric black holes, the validity or violation of the second law in the extended phase space was found to be dependent  of the spin parameter,  radius of the AdS spacetime, and their variations.

 In this paper, we intend to investigate the thermodynamics and weak cosmic censorship conjecture with pressure and volume in the Gauss-Bonnet AdS black hole.     In the extended phase space, it has been found that the  Gauss-Bonnet coefficient  should be treated as
a dynamical variable besides the cosmological parameter in order to  satisfy the Smarr relation \cite{Cai:2013qga}. Thus in the Gauss-Bonnet gravity, the thermodynamic phase space is more extensive than that in \cite{Gwak:2017kkt}.  In this paper, we  want to explore how  the  Gauss-Bonnet coefficient affects the  laws of thermodynamics  and  weak cosmic censorship conjecture besides the
pressure and volume. In addition, in \cite{Gwak:2017kkt},  the
weak cosmic censorship conjecture was fond to be  valid  in the extended phase space for the near-extremal  Reissner Nordstr\"{o}m-AdS  black hole. However, we found that there were some approximations. In our paper, we want to explore whether the weak cosmic censorship conjecture is valid in the extended phase space without approximations.
 As a result, we find the weak cosmic censorship conjecture for the near-extremal  Gauss-Bonnet AdS black hole is  violable, depending on the values of  $\alpha, l, r_{min}$ and their variations.

The remainder of this article is organized as follows. In section 2, we will  briefly review the thermodynamics of the   Gauss-Bonnet AdS   black hole.  In section 3, we are going to get the relation between the energy and momentum of the   absorbed particle near the horizon.  The laws of  thermodynamics and weak cosmic censorship conjecture  will be discussed in the normal and extended phase space in section 4 and  section 5,   respectively. We employ the variation of entropy to check the second law of thermodynamics. We adopt the variation of the minimum value of the function which determine the locations of the horizons  to check the weak cosmic censorship conjecture.  Section 6 is devoted to our conclusions. Throughout this paper, we will set $G_d=\hbar=c
=k=1$.

\section{Review of the Gauss-Bonnet AdS black hole }\label{sec2}
The action admitting the d-dimensional Einstein-Maxwell theory with a Gauss-Bonnet term and a cosmological constant term is   \cite{Boulware:1985wk,Cai:2001dz,Wiltshire:1985us,Cvetic:2001bk}
\be
S=\frac{1}{16\pi }\int d^dx\sqrt{-g}\left[R-2\Lambda +\alpha _{\text{GB}}\left(R_{\mu \nu \gamma \delta }R^{\mu \nu \gamma \delta }-4R_{\mu \nu }R^{\mu \nu }+R^2\right)-4\text{$\pi $F}_{\mu \nu }F^{\mu \nu }\right],\label{1action}
\ee
where $\alpha _{GB}$ is the Gauss-Bonnet coefficient, $\Lambda$ is the cosmological constant that relates to the AdS radius $l$ with the relation $\Lambda =-\frac{(d-1)(d-2)}{2l^2}$, $R$ is the Ricci scalar, $g$ is determinant of the
metric tensor, and $F_{\mu \nu } $ is the Maxwell field strength with the definition  $F_{\mu \nu }=\partial _{\mu }A_{\nu }-\partial _{\nu }A_{\mu }$, where $A_{\mu}$ is the vector potential. The Gauss-Bonnet coefficient is set to be positive for it is proportional to the inverse string tension with positive coefficient in the low energy effective action of heterotic string theory \cite{Boulware:1985wk,Cai:2001dz}.  The dimension of the spacetime is supposed to be larger than four since the Gauss-Bonnet term has  no dynamics in four dimensions. From the action  in  Eq.(\ref{1action}), we can obtain the following solution  \cite{Boulware:1985wk,Cai:2001dz,Wiltshire:1985us,Cvetic:2001bk}
\be
{dS}^2=-f(r){dt}^2+f^{-1}(r){dr}^2+r^2h_{{ij}}{dx}^i{dx}^j, \label{metric2}
\ee
with
\be
f(r)=\kappa +\frac{r^2}{2 \alpha}-\frac{r^2}{2\alpha}\sqrt{1+\frac{64 \pi  \alpha M}{(d-2) \Omega   _{\kappa } r^{d-1}}-\frac{2\alpha Q^2}{ (d-2)(d-3)\text{  }r^{2d-4}}-\frac{4 \alpha}{ l^2}},
\ee
 where $\alpha = (d-3)(d-4)\alpha_{GB}$ is the redefined Gauss-Bonnet coefficient,
  $M$ and  $Q$ are the mass and charge of the black hole respectively, and $r^2h_{{ij}}{dx}^i{dx}^j$   is the line element of a $(d-2)$-dimensional  Einstein space with constant curvature $(d-2)(d-3)\kappa$  and volume $  \Omega   _{\kappa }$.  The value of $\kappa$ can be 1, 0, -1, corresponding to spherical,   flat and hyperbolic topology of black hole horizon. In this paper, we are interested in the case $\kappa=1$, and the corresponded volume of the  $(d-2)$-dimensional  space is labeled as
$  \Omega  $.

The non-vanishing component of the vector potential is
\be
A_t=- \frac{Q r ^{3-d}  \Omega   _{   }}{16 (d-3) \pi }.
\ee
From the equation $f(r)=0$, we can obtain two   solutions, which correspond to the inner horizon and outer horizon. The outer horizon is the event horizon, labelled as $r_h$ thereafter.
 At the event horizon, the mass $M$ can be calculated as
 \be
M=\frac{  \Omega   _{   } r_h^{-d-5} \left(l^2 \left(2 \left(d^2-5 d+6\right)    r_h^{2 d} \left(a +r_h^2\right)+Q^2 r_h^8\right)+2 \left(d^2-5 d+6\right) r_h^{2 d+4}\right)}{32 \pi  (d-3) l^2}.
 \ee
 According to the definition of the surface gravity, the Hawking temperature can be written as
\be
T=\frac{  -l^2 \pi  Q^2 r^8+2 (d-2) \pi  {r_h}^{2 d} \left(\alpha (d-5)    ^2 l^2+(d-3)     l^2 {r_h}^2+(d-1) {r_h}^4\right) }{8 {r_h}^{2d+1}(d-2 )l^2 \left(2 \alpha    +{r_h}^2\right)}.\label{tem}
\ee
The Bekenstein-Hawking entropy, and electric potential  can be obtained by \cite{Cai:2001dz}
\be
S_h=\int _0^{r_h}T^{-1}\left(\frac{\partial M}{\partial {r_h}}\right)_{Q}{d{r_h}}=\frac{{r_h}^{d-4} \left(2 \alpha (d-2)    +(d-4) {r_h}^2\right)  \Omega   _{   }}{4 (d-4) \pi ^2},\label{s}
\ee
\be
\Phi =\left(\frac{\partial M}{\partial Q}\right)_{S_h}=A_t(\infty)-A_t(r_h)=\frac{Q r_h ^{3-d}  \Omega   _{   }}{16 (d-3) \pi },
\ee
in which  we have employed the first law of black hole thermodynamics in the normal phase space, namely the cosmological parameter is fixed. Recent investigations have shown that
the cosmological parameter can be a dynamical quantity, the thermodynamic phase space thus is extended. In the extended phase space, we also can construct the first law of thermodynamics  by treating
 the cosmological
constant  as a dynamical pressure and its conjugate quantity
as the thermodynamic volume. In this case, the black hole mass $M$ is explained as enthalpy $H$ rather than internal energy $U$ of the system.  In addition, as the cosmological constant is regarded  as thermodynamic
pressure in the first law, the Smarr relation for black  hole thermodynamics can be obtained by  scaling argument. In the Gauss-Bonnet gravity, to satisfy the  Smarr relation, the Gauss-Bonnet coefficient also should be treated as a dynamical quantity. The first law in  the extended phase space thus takes the form as \cite{Cai:2013qga,Mahish:2019tgv}
\begin{equation}
{dH}={TdS_h}+{\Phi dQ}+{VdP}+{\mathcal{A}d \alpha},
\end{equation}
 in which $P$ is the pressure, $V$ is its conjugate quantity  interpreted as volume, and  $\mathcal{A}$ is the conjugate quantity of  Gauss-Bonnet coefficient $\alpha$, which are defined respectively as
\be
 P=-\frac{\Lambda }{8\pi }=\frac{(d-1)(d-2)}{16\text{$\pi $l}^2},\label{p}
\ee
\be
V=\left(\frac{\partial H}{\partial P}\right)_{S_h,Q,\alpha}=\frac{r_h ^{d-1}  \Omega   _{   }}{d-1 },\label{v}
\ee
\be
\mathcal{A}=\left(\frac{\partial H}{\partial a}\right)_{S_h,Q,P}=\frac{(d-2)^2 {r_h}^{d-5} \Omega   _{   }}{16 \pi }.\label{a}
\ee
One can check that the following Smarr relation is also satisfied
\be
(d-3)H=(d-2) T S_h-2 P V+2 \mathcal{A} \alpha+(d-3) Q\Phi.
\ee
In the Born-Infeld AdS black hole, the conjugate quantity of Born-Infeld parameter is interpreted as Born-Infeld vacuum polarization \cite{Gunasekaran}. While in the  Gauss-Bonnet AdS black hole, the physical meanings of the conjugate quantity $\mathcal{A}$  is still not known, we only know it has the dimension $[length]^{-3}$ \cite{Cai:2013qga}.

\section{Energy-momentum relation of the absorbed particle}\label{sec3}
 In this section, we intend to obtain the energy-momentum relation of a charged particle near the event horizon as it is absorbed by the Gauss-Bonnet AdS black hole. We are interested in the scalar particle, so we will employ  the following Hamilton-Jacobi equation to study the dynamical of the absorbed particle
\be
g^{\mu \nu }\left(p_{\mu }-\text{eA}_{\mu }\right)\left(p_{\nu }-\text{eA}_{\nu }\right)+\mu ^2=0,  \label{hjequation}
\ee
where
 $p_\mu=\partial_\mu I$
is the momentum, $e$ is the charge, $\mu$ is the mass, and $I$ is the Hamilton action of the particle. In the spherically symmetric spacetimes,  the Hamilton action  of the moving particle can be separated into
\be
I=-{Et}+W(r)+\underset{i=1}{\overset{d-3}{  \Omega   }}I_{\theta_i }(\theta_i )+{L\psi },  \label{action}
\ee
in which  $E$ and $L$ are the energy and angular momentum of the particle respectively,  and the  $(d-2)$-dimensional    sphere  has been expressed  as
\be
h_{{ij}}{dx}^i{dx}^j=\underset{i=1}{\overset{d-2}{\Sigma }}\left(\underset{j=1}{\overset{i}{\Pi }}\sin ^{2}\theta _{j-1}\right)d {\theta_i }^2, \theta_{d-2}\equiv \psi.
\ee
To solve the Hamilton-Jacobi equation,  we will use the inverse metric of the black hole  in Eq.(\ref{metric2})
\begin{align}
g^{\mu \nu }\partial _{\mu }\partial _{\nu }=-f(r)^{-1}\left(\partial _t\right) ^2+f(r)\left(\partial _r\right) ^2+r^{-2}\underset{i=1}{\overset{d-2}{\Sigma }}\left(\underset{j=1}{\overset{i}{\Pi }}\sin ^{-2}\theta _{j-1}\right)\left(\partial _{\theta_i }\right) ^2.  \label{inversemetric}
\end{align}
Substituting Eqs.(\ref{action}) and  (\ref{inversemetric}) into Eq.(\ref{hjequation}), we can obtain
\begin{align}
&-\frac{1}{f(r)}\left(-E-{eA}_t\right) ^2+f(r)\left(\partial _rW(r)\right) ^2+r^{-2}\underset{i=1}{\overset{d-3}{\Sigma }}\left(\underset{j=1}{\overset{i}{\Pi }}\sin ^{-2}\theta _{j-1}\right)\left(\partial _{\theta_i }I(\theta_i )\right) ^2\nonumber\\
&~~~~~+r^{-2}\left(\underset{j=1}{\overset{d-2}{\Pi }}\sin ^{-2}\theta _{j-1}\right)L^2+u^2=0.  \label{eq3.5}
\end{align}
With a variable $\mathcal{K}$ to separate this equation, we can get the radial equation and angular
equation
\begin{align}
-\frac{r^2}{f (r)} \left(-E -\text{eA}_t\right) ^2+r^2f (r) \left(\partial _rW(r)\right) ^2+r^2\mu ^2=-\mathcal{K},  \label{eq3.6}
\end{align}
\begin{align}
\underset{i=1}{\overset{d-3}{\Sigma }}\left(\underset{j=1}{\overset{i}{\Pi }}\sin ^{-2}\theta _{j-1}\right)\left(\partial _{\theta_i }I(\theta_i )\right) ^2+\left(\underset{j=1}{\overset{d-2}{\Pi }}\sin ^{-2}\theta _{j-1}\right)L^2=\mathcal{K}.  \label{eq3.7}
\end{align}
Lastly,  we obtain the radial momentum
\begin{align}
p^r\equiv g^{rr}\partial_r W(r)=f(r)\sqrt{\frac{-\mu ^2r^2+\mathcal{K}}{r^2f(r)}+\frac{1}{f(r)^2}\left(-E -\text{eA}_t\right) ^2}.\label{pr}
\end{align}
As a particle drops into the black hole, we will pay attention to the thermodynamics of the black hole. Therefore, we are interested only the near horizon behavior of the  particle, where $f(r_h)=0$.  Near the event horizon, Eq.(\ref{pr}) will be simplified as
\begin{align}
E=-A(r_h)e+\left|p^r_h\right|,  \label{prh}
\end{align}
in which $p^r_h=p^r(r_h)$. For the $|p^r_h|$ term, we will choose the positive sign thereafter as done in \cite{ref8,Zeng:2019baw,He:2019fti} in order to assure  the particle drops into the black hole in a  positive flow of time direction.

\section{Thermodynamics and  weak cosmic censorship conjecture in the normal phase space}
\label{sec4}

In this section, we will investigate the thermodynamics and  weak cosmic censorship conjecture in the normal phase space on the basis of Eq.(\ref{prh}). We want to explore whether  the energy-momentum relation can check the validity of the second law of thermodynamics and weak cosmic censorship conjecture besides produce the first law of thermodynamics.

\subsection{The first law of thermodynamics in the normal phase space}

In the normal phase space,    the cosmological parameter is fixed, and the black hole is characterized by the mass $M$ and   charge $Q$. The mass $M$ is interpreted as  the internal energy of the thermodynamic system.  As a  particle is absorbed by the black hole, we suppose the energy and charge are conserved, that is, the change of the internal energy and charge of the black hole  satisfy
 \begin{align}
E=dM,   e=dQ.  \label{eq4.1}
\end{align}
In this case, Eq.(\ref{prh}) changes into
\begin{align}
dM=\frac{Q r_h ^{3-d}  \Omega   _{   }}{16 (d-3) \pi }{dQ}+ p^r_h.  \label{eq4.2}
\end{align}
The absorbed charged particle will also change the location of the event horizon. We label the final state of the event horizon as $r_h+dr_h$, which satisfies $f(r_h+dr_h)=0$ too.  According to $f(r_h+dr_h)=f(r_h)=0$,  there is always a relation
\begin{align}
df_h=\frac{\partial f_h}{\partial M}{dM}+\frac{\partial f_h}{\partial Q}{dQ}+\frac{\partial f_h}{\partial r_h}{dr}_h=0.  \label{eq4.3}
\end{align}
Inserting Eq.(\ref{eq4.2}) into Eq.(\ref{eq4.3}), we find both $dM$  and $dQ$ are eliminated.  The solution of $dr_h$ thus can be expressed as
\begin{align}
dr_h=\frac{32 l^2 \pi   \Omega   _{   }^{-1} p^r_hr_h ^{6+d}}{ 2 \left(2-3 d+d^2\right) r_h ^{4+2 d}+l^2 \left(-Q^2 r_h ^8+2 (d-2)     r_h ^{2 d} \left(\alpha (d-5)    +(d-3) r_h ^2\right)\right)  }.  \label{drnormal}
\end{align}
In addition, based on Eq.(\ref{s}), the variation of the entropy can be expressed as
\begin{align}
{dS_h}=\frac{(d-2){  }{r_h}^{d-5} \left(2 \alpha    +{r_h}^2\right)   \Omega   _{   }}{4 \pi ^2}{dr_h}.  \label{vs}
\end{align}
Substituting Eq.(\ref{drnormal}) into  Eq.(\ref{vs}), we have
\begin{align}
{dS_h}=\frac{8 \pi^{-1}  (d-2) l^2  p^r_h r_h ^{1+2 d} \left(2 \alpha     +r_h ^2\right)}{ 2 \left(2-3 d+d^2\right) r_h ^{4+2 d}+l^2 \left(-Q^2 r_h ^8+2 (d-2)    {  }r_h ^{2 d} \left(\alpha (d-5)    +(d-3) r_h ^2\right)\right)}.  \label{vsf}
\end{align}
From Eqs.(\ref{tem}) and (\ref{vsf}), we get
\begin{align}
T {dS_h}=p^r_+.  \label{tspr}
\end{align}
Combining Eqs.(\ref{eq4.2})  and (\ref{tspr}), we find
\begin{align}
{dM}=T {dS_h}+\Phi {dQ}, \label{eq4.7}
\end{align}
which is nothing but the first law  of thermodynamics in the normal phase space. That is, the first law  is valid in the normal phase space as a charged particle is absorbed  by the  Gauss-Bonnet AdS black hole.

\subsection{The second  law of thermodynamics in the normal phase space}
The second law of black hole thermodynamics states that the entropy of the black holes never decrease in the clockwise direction. As a particle is absorbed by the Gauss-Bonnet AdS black hole, the entropy of the black hole also should increase if the second law is valid, namely the variation of the entropy should satisfies $dS_h>0$. In this section, we will employ
Eq.(\ref{vsf}) to check whether this is true.

For an extremal black hole, its temperature vanishes. From Eq.(\ref{tspr}), we know that the variation of entropy  is divergent, which is meaningless. Thus we mainly concentrate on the
near-extremal black hole thereafter.  We will obtain $dS_h$ by numeric method. During the numeric calculation, we focus on studying how $\alpha$ and $d$ affect the value of $dS_h$  by fixing $Q= 7$,  $l=p^r=  \Omega   =1$. For a given values of  $\alpha$ and $d$, we can obtain the mass of the extremal black hole by solving equation $f(r_h)=0$. For the case that the two roots are the same, the corresponded mass is  that of the extremal black hole. For example, when $d=5$, the masses of the extremal black holes are  0.478672011, 0.41898843, 0.36527364 for $\alpha=2, 1,  0.1$ respectively, which is shown in Table 1. The mass of the non-extremal black holes should be larger than that of the extremal black hole.  For different $d$ and $\alpha$, the values of  $dS_h$ are given in  Table 1, Table 2, and Table 3.  From these tables, we can see clearly  that  the variation of the entropy is positive always for the extremal and non-extremal black holes. For the extremal black holes, though the values seem finite, there are much lager than the near extremal black hole, which can be regarded as divergent. So, the second law of thermodynamics is valid for all the black holes  in the normal phase space.

In addition, from  Table 1, Table 2, and Table 3, we also can observe how  $d$ and $\alpha$
affect the values  of     $dS_h$. For a fixed  values of  $d$ and $\alpha$, we find the values of
  $dS_h$ decrease
as the black holes move far away from the extremal black hole.  As the value
 of $d$ is fixed, the value of the mass
of the extremal black hole   decreases as the value of $\alpha$ reduces. As the
mass of a black hole  is fixed,  the value of  $dS_h$ decreases while the value of
   $r_h$   increases as $\alpha$ reduces. When the value of $\alpha$ is fixed, we also can see how $d$ affect the value of $dS_h$ and  $r_h$. As
  the value of  $d$ increases,
    the value of the  mass of   extremal black hole decreases, while  $dS_h$  and
   $r_h$   increases.

\begin{center}
{\footnotesize{\bf Table 1.} The relation between $ {dS}_h$, $M$ and $r_h$  for $d=5$ in the normal phase space.\\
\vspace{2mm}
\begin{tabular}{ccc|ccc|ccc}
\hline
 $\alpha=2$     &             &     &  $\alpha=1$ & & &  $\alpha=0.1$ &  \\ \hline
$M$       &$r_h $  & $dS_h $&  $M$ &$r_h $& $dS_h $  &  $M$ &$r_h $& $dS_h $  \\
\hline
0.478672011  & 1.059899   & $275.84 $ & 0.41898843 &1.05925 &4004.7 & 0.36527364 &1.05928 &825.02  \\
0.4787     & 1.064454   & $35.889$  & 0.4190 &1.06256 &34.250 &0.3653 & 1.06426 &9.6753  \\
0.48    & 1.095201         & $5.3898 $  & 0.42 &1.09059 &3.7818 & 0.370& 1.12751&0.8177   \\
0.5     & 1.206054         & $1.4548$   & 0.5 & 1.34925 &0.5288 & 0.5& 1.43458&0.2413  \\
0.6    & 1.415228  & $0.6771 $ & 0.6 &1.49469 &0.3878 &0.6&1.55497&0.2086 \\
0.7     & 1.540696     & $0.5243 $ & 0.7 &1.60113 & 0.3296 &0.7&1.64970&0.1914 \\
0.8       &1.638024         & $0.4479$   & 0.8 &1.68814 & 0.2951 &0.8&1.72953&0.1799\\
0.9     &1.719498  &$0.3996$ & 0.9 & 1.76288& 0.2715 &0.9&1.79928&0.1717 \\
\hline
\end{tabular}}
\end{center}
\begin{center}
{\footnotesize{\bf Table 2.}  The relation between $ {dS}_h$, $M$ and $r_h$  for $d=6$ in the normal phase space.\\
\vspace{2mm}
\begin{tabular}{ccc|ccc|ccc}
\hline
$\alpha=2$     &             &     &  $\alpha=1$ & & &  $\alpha=0.1$ &  \\ \hline
$M$       &$r_h $  & $dS_h $&  $M$ &$r_h $& $dS_h $  &  $M$ &$r_h $& $dS_h $  \\
\hline
0.469357085      & 0.927431       & $12640.4 $   &0.39485042&0.94529&1587.9&0.32655982&0.961894&1084.7 \\
0.4694           & 0.931772         & $23.3241$    &0.3949&0.95002&13.440&0.3266&0.96623&6.0210\\
0.47             & 0.944334         & $6.19426 $  &0.40&0.99447&1.4585&0.33&1.00257&0.7310  \\
0.5              & 1.047712         & $1.09073$   &0.5&1.17392&0.4427 &0.5&1.25735&0.1915 \\
0.6              & 1.180504         & $0.63647 $   &0.6&1.26516&0.3576 &0.6&1.33130&0.1744 \\
0.7              & 1.264667         & $0.52439$   &0.7&1.33420&0.3171&0.7&1.39099& 0.1645 \\
0.8             &1.330526         & $0.46491$    &0.8&1.39119&0.2917&0.8&1.44180& 0.1577\\
0.9              &1.385745            &$0.42571$  &0.9&1.44030&0.2735 &0.9&1.48639& 0.1526\\
\hline
\end{tabular}}
\end{center}
\begin{center}
{\footnotesize{\bf Table 3.}  The relation between $ {dS}_h$, $M$ and $r_h$  for $d=7$ in the normal phase space.\\
\vspace{2mm}
\begin{tabular}{ccc|ccc|ccc}
\hline
$\alpha=2$     &             &     &  $\alpha=1$ & & &  $\alpha=0.1$ &  \\ \hline
$M$       &$r_h $  & $dS_h $&  $M$ &$r_h $& $dS_h $  &  $M$ &$r_h $& $dS_h $  \\
\hline
0.463091855      & 0.878523         & $1642.9 $ &0.38442131&0.90054&1129.2&0.31006521&0.92257& 294.88  \\
0.4631            & 0.879883         & $40.076$   &0.3845&0.90504& 8.4265&0.3101&0.92564&5.1496\\
0.47              & 0.919834         & $1.5810 $ &0.40&0.96557& 0.7360&0.32&0.97599&0.3784 \\
0.5              & 0.975948        & $0.8116$   &0.5&1.08097&0.3697 &0.5&1.15811&0.1586\\
0.6              & 1.069715         & $0.5323 $  &0.6&1.14703&0.3125 &0.6&1.21152&0.1475  \\
0.7              & 1.131048         & $0.4581 $  &0.7&1.19760&0.2843&0.7&1.25492&0.1409   \\
0.8             &1.179453          & $0.4183$   &0.8&1.23951&0.2663 &0.8&1.29194&0.1364 \\
0.9              &1.220202          &$0.3917$  &0.9&1.27567&0.2533&0.9&1.32443&0.1330 \\
\hline
\end{tabular}}
\end{center}

\subsection{Weak  cosmic censorship conjecture in the normal phase space}

The weak
cosmic censorship conjecture  states that an  observer located at future null infinity can not observe the singularity of a spacetime  for the singularity is hidden by the horizon.
In this section, we want to check whether this is true as a particle drops into the Gauss-Bonnet black hole. We will explore whether there is a horizon after the particle is absorbed.

 The event horizon of the black hole is determined by the function $f(r)$, so we concentrate on studying  how $f(r)$ changes. The function $f(r)$ has a minimum value at radial coordinate $r_\text{min}$.    For the case $f(r_\text{min })<0$,  there are two roots, for the case $f(r_\text{min })=0$, the two roots coincide, and the black hole becomes into an extremal black hole,  for the case $f(r_\text{min })>0$, the function  has no real root so that there is not an event horizon. Our motivation is to explore  how  $f(r_\text{min})$  moves  as a charged particle is absorbed by the Gauss-Bonnet AdS black hole. For the initial state is a black hole, $f(r_\text{min })$ satisfies the following conditions \cite{Gwak:2019asi,Gwak:2015fsa,Gwak:2018tmy,Chen:2019pdj,Chen:2019nsr,Hong:2019yiz}
\be
f(M,Q,\alpha, l,r)|_{r=r_\text{min}}\equiv f_\text{min}=\delta\leq 0,\label{condition1}
\ee
\be
\partial_{r} f(M,Q,\alpha,l,r)|_{r=r_\text{min}}\equiv f'_\text{min}=0,  \label{condition2}
\ee
\be
 (\partial_{r})^2 f(M,Q,\alpha,l,r)|_{r=r_\text{min}}\equiv f''_\text{min}>0.  \label{condition3}
\ee
 For the extremal black hole, $\delta=0$, $r_h$ and $r_\text{min }$ are coincident. For the near-extremal black hole, $\delta$ is a small quantity, $r_\text{min }$ locates at the middle of the two horizons.

In the normal phase space, the state parameters of the black hole are the mass $M$ and charge $Q$. As a charged particle is absorbed, the mass and charge of the black hole will change into $M+dM$ and $Q+dQ$. Correspondingly, the horizon $r_h$ and radial coordinate $r_\text{min }$ will change into $r_h+d r_h$ and  $r_\text{min }+dr_\text{min }$.
Note that  Eq.(\ref{condition2}) is satisfied at both $r_\text{min }$ and $r_\text{min }+{dr}_\text{min }$, which implies
\be
{df}'_\text{min }=\frac{\partial f'_\text{min }}{\partial M} dM+\frac{\partial f'_\text{min }}{\partial Q} dQ+\frac{\partial f'_\text{min }}{\partial r_\text{min }} dr_\text{min }=0.  \label{eq4.9}
\ee
In addition, at $r_\text{min }+{dr}_\text{min }$, the function $ f\left(r_\text{min }+dr_\text{min }\right)$ can be expressed as
\be
f\left(r_\text{min }+dr_\text{min }\right)=f_\text{min }+df_\text{min }=\delta+\frac{\partial f_\text{min }}{\partial M} dM+\frac{\partial f_\text{min }}{\partial Q} dQ. \label{eq4.11}
\ee
Firstly, we are interested in the extremal black hole, where $f_\text{min }=\delta =0$ and  Eq.(\ref{eq4.2}) is applicable.  Inserting Eq.(\ref{eq4.2}) into Eq.(\ref{eq4.11}), we can get finally
\be
df_{\min }=\text{  }-\frac{16 \pi   p^r_hr_{\min } ^{-(d-5)}}{(d-2) \left(2 \alpha    +r_{\min } ^2\right)   \Omega   _{   }}, \label{eq4.12}
\ee
 which seems to be  negative always. However, note that the black hole is an  extremal black hole here, from Eq.(\ref{tspr}), we know that $  p^r_h=0$, so $df_\text{min }=0$, implying that the extremal Gauss-Bonnet AdS  black hole is stable. In other words,  as a particle drops into the extremal Gauss-Bonnet AdS black hole,  the final state of the black hole is still an extremal black hole. The weak cosmic censorship conjecture thus is valid  in this case.

 \begin{figure}[h]
\centering
\includegraphics[trim=0cm 0cm 10cm 20.5cm, clip=true, scale=1]{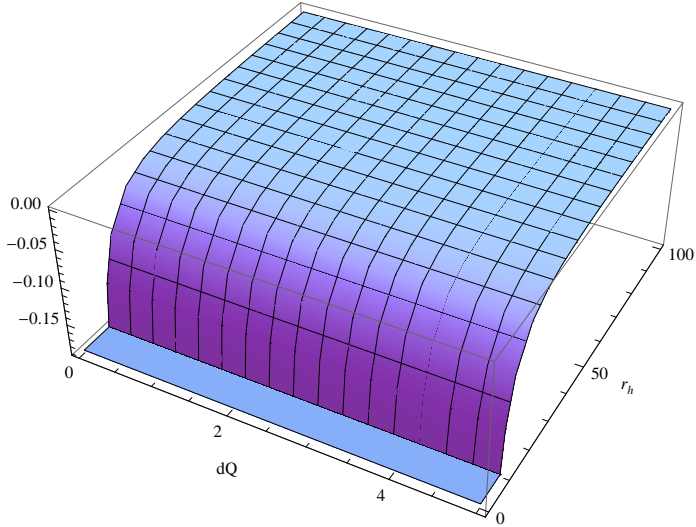}
\caption{The  value of ${df}_{\min }$  for different $dQ$ and $r_h$ for  $p^r = \alpha = \Omega    = 1$, $ \epsilon = 0.0001$, $d=5$.}
\label{fig66}
\end{figure}

For a near-extremal black hole, Eq.(\ref{eq4.2}) is not applicable since $r_m$ and $r_h$ are not coincident. With the condition $r_h=r_\text{min }+\epsilon $, we can expand Eq.(\ref{eq4.2}) at $r_\text{min}$, which leads to
\be
\text{dM}=\left( p^r_h+\frac{ Q r_{\min } ^{3-d}   \Omega   _{   } \text{dQ}}{16 (d-3) \pi }\right)-\frac{\left(Q r_{\min } ^{2-d}   \Omega   _{   } \text{dQ}\right)\epsilon }{16 \pi }+O(\epsilon )^2. \label{eq4.13}
\ee
Combining  Eq.(\ref{eq4.11}) and Eq.(\ref{eq4.13}),  we have
\be
{df}_{\min }=-\frac{16 \left(\pi   p^r_h r_{\min } ^{5-d}\right)}{(d-2) \left(2 \alpha    +r_{\min } ^2\right)   \Omega   _{   }}+\frac{ Q r_{\min } ^{7-2 d} \epsilon  \text{dQ}}{(d-2) \left(2 \alpha    +r_{\min } ^2\right)} +O(\epsilon)^2. \label{eq4.14}
\ee
In the Eq.(\ref{eq4.14}), $\epsilon$ is a very small quantity, so the  third terms can be neglected approximately. In addition, comparing with the first term, the second term is smaller than it always. In this case, Eq.(\ref{eq4.14})  is negative too, please refer to Figure (\ref{fig66}). Therefore, for the near-extremal black hole, the weak cosmic censorship conjecture is also valid under a charged particle absorption in the normal phase space.

In fact, the high order corrections  are important to discuss the  weak cosmic censorship conjecture \cite{Wald2018}. However, in our paper,   we find it has  little effect, which is shown in Figure \ref{fig66}. From this figure, we know that for different $dQ$ and $r_h$, the value of ${df}_{\min }$ is negative always.  As the values of $p^r, \alpha,   \Omega  , \epsilon, d$ change, it is still negative too though the values of  ${df}_{\min}$ changes, which is not shown here.

\section{Thermodynamics and weak cosmic censorship conjecture in the extended phase space}\label{sec5}

In the normal phase space, we have derived the first law of thermodynamics and found that the second law as well as the weak cosmic censorship conjecture are valid. In this section, we intend to discuss the thermodynamics and weak cosmic censorship conjecture in the extended phase space, where the pressure and volume emerge. We will explore whether the first law, second law, as well as the weak cosmic censorship conjecture are valid in this framework.

\subsection{The first law of thermodynamics in the extended phase space}

In the extended phase space, the mass $M$ is not the internal energy $U$ but the enthalpy $H$ of the thermodynamic system. The relation between the internal energy and enthalpy is \cite{Cai:2013qga}
\be
M=U+PV+\alpha\mathcal{A}.  \label{eq3.13}
\ee
In this case, the variation of the energy and charge  takes the form  as
\be
E=dU=d(M-PV-\alpha\mathcal{A}),  e=dQ. \label{eq3.14}
\ee
Correspondingly, the relation between the energy and momentum in Eq.(\ref{prh}) should be  rewritten as
\be
dU=d(M-PV-\alpha\mathcal{A})=\frac{Q r_h ^{3-d}  \Omega   _{   }}{16 (d-3) \pi }{dQ}+ p^r_h. \label{epspr}
\ee
In addition, in order to  obtain the first law of thermodynamics in the extended phase space,  besides $dS$, we also should find   $dV$ and $d \mathcal{A}$. From Eq. (\ref{v}) and Eq. (\ref{a}), we have
\begin{align}
dV=  \Omega   r^{d-2}_h dr_h,\label{dv}
\end{align}
\be
d \mathcal{A}=\frac{(d-5) (d-2)    \Omega   r_h^{d-6}}{16 \pi } dr_h. \label{da}
\ee
To get the final results of $dV$ and $d \mathcal{A}$, we should find  $dr_h$.

In the extended phase space, the state parameters of the Gauss-Bonnet AdS black hole are $M, Q, l, \alpha$, as a particle drops into the black hole,  the state parameters change into $M+dM, Q+dQ, l+dl, \alpha+d\alpha$. Correspondingly, the event horizon will changes into $r_h+dr_h$. Based on the fact that  $f(r_h+dr_h)=f(r_h)=0$, we find
\be
d{f_h}=\frac{\partial f_h}{\partial M}{dM}+\frac{\partial f_h}{\partial Q}{dQ}+\frac{\partial f_h}{\partial l}{dl}+\frac{\partial f_h}{\partial r_h}{dr}_h+\frac{\partial f_h}{\partial \alpha}d\alpha=0.  \label{eq3.17}
\ee
Combining Eqs (\ref{eq3.17}) and (\ref{epspr}), we find    all the variables are eliminated except for $dr_h$ and $p^r_h$, so we get
\be
{dr}_h=-\frac{32 \pi   p^r_h r_h ^{4+d}}{\left(Q^2 r_h ^6-12     r_h ^{2 d}+10 d     r_h ^{2 d}-2 d^2 {   r}_h ^{2 d}\right)   \Omega   _{   }}.  \label{eq3.20}
\ee
Inserting Eq.(\ref{eq3.20}) into Eqs.(\ref{vs}), (\ref{dv}), and (\ref{da}), we get
\be
{dS}_h=\frac{8 (d-2)  p^r_h\left(2 \alpha    +r_h ^2\right)}{\pi  r_h \left(2 \left(6-5 d+d^2\right)    -Q^2 r_h ^{6-2 d}\right)},  \label{dsf}
\ee
\be
{dV}=\frac{32 \pi   p^r_h r_h ^{2+2 d}}{-Q^2 r_h ^6+2 \left(6-5 d+d^2\right) r_h ^{2 d}},   \label{dvf}
\ee
\be
d \mathcal{A}=
\frac{2 (d-5) (d-2)   p^r_h}{r_h^2 \left(2 \left(d^2-5 d+6\right) -Q^2 r_h^{6-2 d}\right)}.\label{ddaf}
\ee
Combining Eqs (\ref{tem}), (\ref{p}),  (\ref{dsf}), (\ref{dvf}), (\ref{ddaf}),  we find
\begin{align}
{TdS_h}-{PdV}-{\alpha d\mathcal{A}}= p^r_h.  \label{epr}
\end{align}
Substituting Eq (\ref{epr}) into  Eq (\ref{epspr}), we get
\begin{align}
dU=\Phi dQ + T dS_h-PdV-\alpha d \mathcal{A}.\label{in}
\end{align}
The relation between the  internal energy and enthalpy in  Eq (\ref{eq3.13}) also can be written  as
\be
dM=dU+PdV+VdP+\alpha d\mathcal{A}+\mathcal{A} d\alpha.  \label{ihr}
\ee
Substituting Eq (\ref{ihr}) into  Eq (\ref{in}), we can obtain lastly
\begin{align}
dM=T dS_h+\Phi  dQ+V  dP+\mathcal{A} d\alpha,  \label{efirstlaw}
\end{align}
 which is the first law of thermodynamics in the extended phase space. That is, the first law of the Gauss-Bonnet AdS black hole
  in the extended phase space can be obtained under a charged particle absorbtion.
\subsection{The second  law of thermodynamics in the extended phase space}
In the extended phase space, we have  proved that the first law of thermodynamics is  valid.
However, the validity of the first law does not means the second law is valid \cite{Gwak:2017kkt}.
So we should  check the second law in the extended phase space. The second law of thermodynamics states that the entropy of the black hole never decreases.
As the particle is absorbed, the entropy of the final state  thus should be larger than the initial state according to the second law of thermodynamics. Next, we will  check whether this is true with Eq.(\ref{dsf}).

We first study the case of   the extremal black hole, for which  the temperature vanishes. On the basis of Eq.(\ref{tem}), we can obtain a critical charge
\be
Q_c=\frac{\sqrt{2} \sqrt{d-2} r_h^{d-4} \sqrt{\alpha d l^2-5 \alpha l^2+d l^2 r_h^2+d r_h^4-3 l^2 r_h^2-r_h^4}}{l}. \label{cq}
\ee
 Substituting Eq.(\ref{cq}) into Eq.(\ref{dsf}), we   get finally
\begin{align}
{dS}_h=-\frac{4 l^2 p^r_hr_h \left(2 \alpha    +r_h ^2\right)}{\pi  \left(\alpha (d-5)    ^2 l^2+(d-1) r_h ^4\right)},  \label{eq3.26}
\end{align}
which is negative, implying that the second law is invalid for the extremal Gauss-Bonnet AdS black hole.

Next, we focus on investigating the non-extremal black hole.  We will adopt the same numerical method as that in the normal phase space. The difference is that we will employ  Eq.(\ref{dsf}) to obtain the variance of entropy. We also set $ \Omega=l = p^r = 1$, and $Q=7$.  For given values of  $d$ and $\alpha$, we can obtain the masses of the extremal black holes. For example, for $d=6$, the masses of extremal black holes are  0.469357085,
0.39485042, 0.32655982 for $\alpha=2, 1, 0.1$ respectively. For any non-extremal black holes with different $d$ and $\alpha$,   we can calculate the values of $r_h$ and $dS_h$,  which are listed in Table 4, Table 5, and Table 6. From these tables, we find that as the
mass  of the black hole increases, the event horizon of the black hole increases.   While for $dS_h$, there is a divergent point,   which divides the variation of entropy into positive region and negative region.  The variation of entropy is negative for the near-extremal  black holes while positive for the far-extremal black holes. That is, the second law of thermodynamics is violated for the extremal Gauss-Bonnet AdS black hole in extended phase space. This conclusion is independent of the values of  $d$ and $\alpha$.

\begin{center}
{\footnotesize{\bf Table 4.} The relation between $ {dS}_h$, $M$ and $r_h$  for $d=5$ in the extended phase space.\\
\vspace{2mm}
\begin{tabular}{ccc|ccc|ccc}
\hline
$\alpha=2$     &             &     &  $\alpha=1$ & & &  $\alpha=0.1$ &  \\ \hline
$M$       &$r_h $  & $dS_h $&  $M$ &$r_h $& $dS_h $  &  $M$ &$r_h $& $dS_h $  \\
\hline
0.478672011     & 1.059899           & $-1.377$  &0.41898843&1.05925& -0.8363&0.36527364&1.05928&   -0.3542                                             \\
0.4787         & 1.064454           & $-1.408$    &0.4190&1.06256& -0.8509&0.3653&1.06426& -0.3652                                               \\
0.48              & 1.095201           & $-1.644 $  &0.42&1.09059& -0.9869  &0.37&1.12751&-0.5442                                                \\
0.50            & 1.206054           & $-3.096$   &0.5&1.34925& -7.7667  &0.45&1.35594&-4.6024                                                 \\
0.60            & 1.415228           & $-150.8$  &0.5315&1.40189&   -31.478   &0.4865&1.41508& -54.04                                           \\
0.60125              & 1.417078       & $-214.2 $  &0.5485&1.42717& 114.439 &0.4925&1.42388& 150.74                                             \\
0.605             & 1.422573          & $913.54 $  &0.55&1.42932&83.1982   &0.5&1.43458&27.901                                              \\
0.65               & 1.482854          & $17.119 $  &0.6&1.49470&9.9134   &0.6&1.55497& 3.5542                                      \\
0.70               & 1.540696          & $9.5658$   &0.7&1.60113&4.79158 &0.7&1.64970&2.5127                                        \\
0.80               &1.638024           &$6.0014$   &0.8&1.68814&3.67834   &0.8&1.72953&2.1607                                   \\
0.90                &1.719498      &$4.8332$   &0.9&1.76288&3.19564 &0.9&1.79928& 1.9925                                           \\
\hline
\end{tabular}}
\end{center}
\begin{center}
{\footnotesize{\bf Table 5.} The relation between $ {dS}_h$, $M$ and $r_h$  for $d=6$ in the extended phase space..\\
\vspace{2mm}
\begin{tabular}{ccc|ccc|ccc}
\hline
$\alpha=2$     &             &     &  $\alpha=1$ & & &  $\alpha=0.1$ &  \\ \hline
$M$       &$r_h $  & $dS_h $&  $M$ &$r_h $& $dS_h $  &  $M$ &$r_h $& $dS_h $  \\
\hline
0.469357085     & 0.927431           & $-1.007$  &0.39485042&0.945294& -0.698&0.32655982&0.96189&-0.315                              \\
0.4694       & 0.931772          & $-1.046$       &0.3949&0.950017& -0.729&0.3266&0.96623&-0.331                        \\
0.47              & 0.944334           & $-1.170 $ &0.40&0.994473&-1.149 &0.35&1.06945&-1.463                               \\
0.50            & 1.047712          & $-3.799$&0.450&1.10992& -13.43   &0.38&1.12536&-107.1                                 \\
0.5475            & 1.122003        & $-85.22$ &0.4515&1.12219& -55.23&0.381&1.12691&178.29                               \\
0.5525              & 1.128293           & $190.46 $ &0.4615&1.12662& 787.30  &0.395&1.14719&5.3786                                \\
0.56           &1.137380     &$34.707$ &0.5 &1.17392 &5.5541 &0.50 &1.25735 &1.2439                                                      \\
0.60           & 1.180504         & $7.8941$   &0.60&1.26516&2.4055 &0.60&1.33130& 0.9929                                                \\
0.70               & 1.264667         & $3.7510$ &0.70&1.33420&1.8846 &0.70&1.39099&0.9070                                          \\
0.80               &1.330526           &$2.9123$   &0.80&1.39119&1.6712 &0.80&1.44180&0.8681                                       \\
0.90                &1.385746      &$2.5478$ &0.90&1.44030&1.5566    &0.90&1.48639&0.8486                                               \\
\hline
\end{tabular}}
\end{center}
\begin{center}
{\footnotesize{\bf Table 6.}  The relation between $ {dS}_h$, $M$ and $r_h$  for $d=7$ in the extended phase space.\\
\vspace{2mm}
\begin{tabular}{ccc|ccc|ccc}
\hline
$\alpha=2$     &             &     &  $\alpha=1$ & & &  $\alpha=0.1$ &  \\ \hline
$M$       &$r_h $  & $dS_h $&  $M$ &$r_h $& $dS_h $  &  $M$ &$r_h $& $dS_h $  \\
\hline
0.463091855     & 0.878523          & $-0.705$  &0.38442131&0.900537&-0.542&0.31006521&0.922567& -0.272                                                \\
0.4631        & 0.879883         & $-0.717$     &0.3845&0.905037&-0.576&0.3101&0.925639&-0.286                                               \\
0.47             & 0.919834          & $-1.206 $&0.39&0.939131&-0.953&0.34&1.01595& -4.866                                                   \\
0.50            & 0.975948           & $-3.307$        &0.40&0.965567&-1.556&0.3515&1.03265& 7.4142                                              \\
0.545            & 1.025403           & $-694.5$   &0.435&1.01910&-17.93&0.3715&1.05687&1.8621                                                 \\
0.548          &1.028150              &$82.552$      &0.44& 1.02494&-160.9&0.4&1.08520&1.1128                                                   \\
0.55             & 1.029951          & $47.969 $     &0.441&1.02607&321.84&0.5&1.15810& 0.6817                                         \\
0.60            & 1.069715        & $5.3613 $      &0.6&1.14703&1.5564&0.6&1.21152&0.5953                                              \\
0.70               & 1.131048          & $2.7381$  &0.7&1.19760&1.2847&0.7&1.25492&0.5621                                         \\
0.80               &1.1794532           &$2.1622$  &0.8&1.23951&1.1641&0.8&1.29194&0.5468                                         \\
0.90                &1.220203      &$1.9073$      &0.9&1.27567&1.0967&0.9&1.32443&0.5395                                              \\
\hline
\end{tabular}}
\end{center}

     We also can investigate how $d$ and $\alpha$ affect the values of  $dS_h$. From Table 4, Table 5, and Table 6, we find that as the values of $\alpha$ decrease, the values of the critical  horizon where $dS_h$ is divergent become   smaller. And as the values of $d$ decrease, the values of the  divergent point become   smaller too.

\begin{figure}[h]
\centering
\includegraphics[trim=0cm 0cm 10cm 22cm, clip=true,scale=1.0]{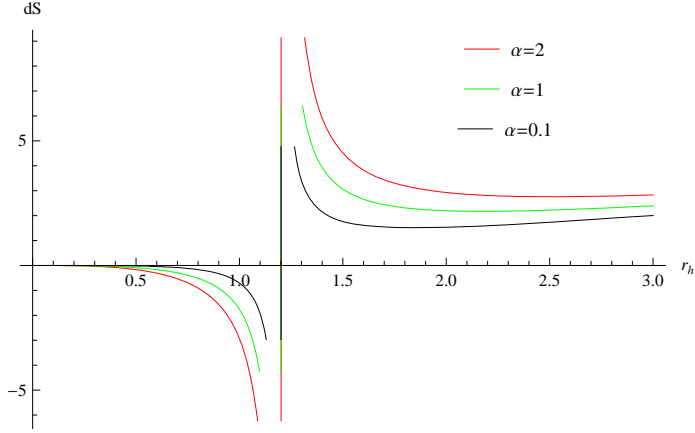}
\caption{The relation between $dS$ and $r_h$ for different $\alpha$ with  $d=5$.}
\label{fig:1}
\end{figure}

In fact, the relation between $dS_h$ and $r_h$ also can be plotted on the basis of Eq.(\ref{dsf}). In Figure 2, Figure 3, and Figure 4, we fix the dimension $d$ to investigate the effect of $\alpha$ on $dS_h$. Obviously, there is a phase transition point which divides $dS_h$ into two branches. Interestedly,  the phase transition point is independent of the values of  $\alpha$.  Taking the case $d=6$ as an example, which is shown in  Table 5.  From  Table 5, we know that the radius of the extremal black holes are 0.927431, 0.945294, 0.96189. While from Figure 3, we know that the phase transition point is about 1.12.
Thus, for the near-extremal black holes, $dS_h$ is negative always, which is independent of $\alpha$. From  Figure 3, we  can conclude that the second law is violated for the near-extremal black holes, which is consistent with that obtained in Table 4.
For the case $d=7$, the phase transition point is about 1.02, which is larger than the radius of the extremal black holes listed in Table 6. Therefore, we  also can conclude that the second law is violated for the near-extremal black holes. This conclusion will not be changed for $d=5$, which is shown in Figure 2. The invalidity of the second law for the near-extremal Gauss-Bonnet AdS black holes thus is universal, independent of the the values of $\alpha$ and  $d$.
\begin{figure}[h]
\centering
\includegraphics[trim=0cm 0cm 10cm 22cm, clip=true, scale=1.0]{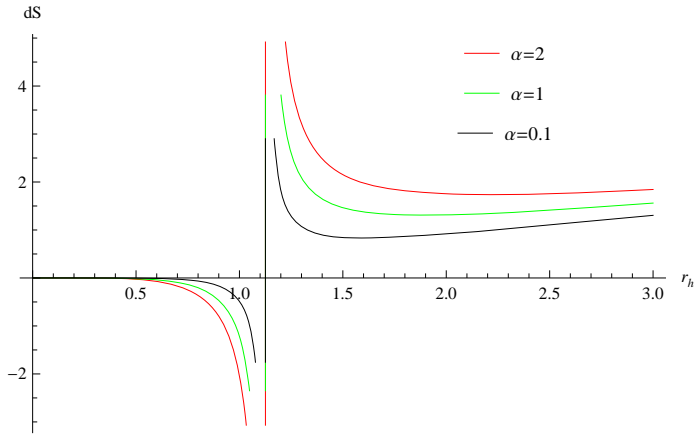}
\caption{The relation between $dS$ and $r_h$ for  different $\alpha$ with  $d=6$.}
\label{fig:2}
\end{figure}

\begin{figure}[h]
\centering
\includegraphics[trim=0cm 0cm 10cm 22cm, clip=true,scale=1.0]{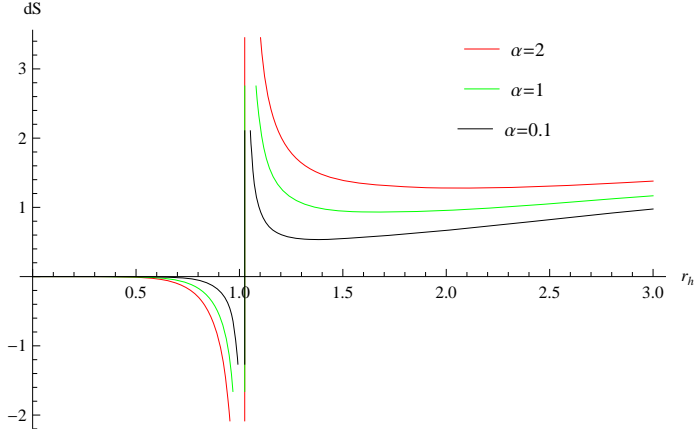}
\caption{The relation between $dS$ and $r_h$ for different $\alpha$ with  $d=7$.}
\label{fig:3}
\end{figure}

\subsection{Weak cosmic censorship conjecture in the extended phase space}

 In the extended phase space,  the second law of thermodynamics was found to be invalid for the extremal and near-extremal black holes, which is different from that in the normal phase space. Therefore, it is necessary to check whether  the weak cosmic censorship conjecture is valid in these cases.

 Similar to that in the normal phase space, we will also investigate how $f(r)$ changes as a charged particle is absorbed. The difference is that in this case, the state parameters are $M, Q,\alpha, l$. The final state of the black hole thus should be the function of $M+dM, Q+dQ,\alpha+d\alpha, l+dl$. Correspondingly, the horizon and minimum point of the final state are $r_h+dr_h$, $r_\text{min}+dr_\text{min}$. At $r_\text{min}+dr_\text{min}$, we find there is always a relation
\be
\partial_{r} f(r)|_{r=r_\text{min}+dr_\text{min}}=f'_\text{min}+df'_\text{min}=0.  \label{frp}
\ee
Using the  condition $f'_\text{min}=0$ in Eq.(\ref{condition2}), we  obtain $df'_\text{min}=0$. Expanding it further, we get
\be
{df}'_{\min }=\frac{\partial f'_{\min }}{\partial M}{dM}+\frac{\partial f'_{\min }}{\partial Q}{dQ}+\frac{\partial f'_{\min }}{\partial l}{dl}+\frac{\partial f'_{\min }}{\partial r_{\min }}{dr}_{\min }+\frac{\partial f'_{\min }}{\partial \alpha}{d\alpha}=0.  \label{dfrp}
\ee
 At   $r_\text{min}+dr_\text{min}$,  the function $f(r)$ takes the form as
\begin{align}
&f|_{r=r_{\min }+{dr}_{\min }}=f_{\min }+{df}_{\min } \nonumber\\
&~~~~~=\delta +\left(\frac{\partial f_{\min }}{\partial M}{dM}+\frac{\partial f_{\min }}{\partial Q}{dQ}+\frac{\partial f_{\min }}{\partial l}{dl}+\frac{\partial f_{\min }}{\partial \alpha}{d\alpha}\right).  \label{eq3.31}
\end{align}
For the extremal black holes, we know  $f'_\text{min}=0$ and $ f_\text{min }=\delta=0 $.  Substituting $dM$ in (\ref{epspr}) into Eq.(\ref{eq3.31}),  we get
\begin{align}
{df}_{\min }=-\frac{r_{\min }^{3-d} \left(16 \alpha {d\mathcal{A}} l^2 \pi  r_{\min } ^2+16 l^2 \pi   p^r_h r_{\min } ^2+\left(2-3 d+d^2\right) {dr}_{\min } r_{\min } ^d \Omega   _{   }\right)}{(-2+d) l^2 \left(2 \alpha    +r_{\min } ^2\right)  \Omega   _{   }}. \label{eq3.32}
\end{align}
Inserting $p^r_h$ in Eq.(\ref{epr}) into Eq.(\ref{eq3.32}), we find
\be
{df}_{\min }=-\frac{ l^2 \left(2 (d-2) r_{\min }^{2 d} \left(\alpha  (d-5)+(d-3) r_{\min }^2\right)-Q^2 r_{\min }^8\right)+2 \left(d^2-3 d+2\right) r_{\min }^{2 d+4}}{2 (d-2) l^2 \left(2 \alpha +r_{\min }^2\right)r_{\min }^{2 d+1}}{dr_{\min }}.\label{dff}
\ee
For the extremal black holes, the extremal charge in Eq.(\ref{cq}) is also applicable. Substituting Eq.(\ref{cq}) into Eq.(\ref{dff}), we gat lastly
\begin{align}
df_{\min }=0,  \label{eq3.33}
\end{align}
which means that $f(r_{\min })$ does not change as a charged particle drops into the extremal Gauss-Bonnet AdS black hole. The extremal black hole thus is still an extremal black hole. The weak cosmic censorship conjecture thus is valid for there is always a horizon to hidden the singularity.

For the near-extremal black hole, Eq.(\ref{epspr})  is not applicable. But we can expand it   near the minimum  point with the relation $r_h=r_\text{min }+\epsilon $.    To the first order,  we find
\begin{align}
{dM}&=\frac{{  }r_{\min } \left( l^3 Q r_{\min } ^8{dQ}+\left(6-5 d+d^2\right)r_{\min } ^{2 d} \left({d\alpha}    ^2 l^3-2 {dl} r_{\min } ^4\right)\right)  \Omega   _{   }}{16 (d-3) l^3 {\pi r}_{\min } ^{6+d}} \nonumber\\
&+\frac{(d-3) \left(2-3 d+d^2\right){  }r_{\min } ^{4+2 d}{dr}_{\min }  \Omega   _{   }}{16(d-3) l \pi } \nonumber\\
&+\frac{(d-3){   }\left(-Q^2 r_{\min } ^8+2 (d-2)     r_{\min } ^{2 d} \left(\alpha (d-5)    +(d-3) r_{\min } ^2\right)\right)  \Omega   _{   }}{32 (d-3)\text{  }\pi }\nonumber\\
&-\frac{ r_{\min } \left( l^3 {Qr}_{\min } ^8{dQ}-(d-2) r_{\min } ^{2 d} \left((d-5) {d\alpha}    ^2 l^3-2 (d-1) {dl} r_{\min } ^4\right)\right)  \Omega   _{   } \epsilon }{16 l^3 {\pi r}_{\min } ^{d+7}}\nonumber\\
&+\frac{(d-2){  }\left(2 \left(2-3 d+d^2\right) r_{\min } ^{4+2 d}{dr}_{\min }\right)  \Omega   _{   } \epsilon }{32 l^2 {\pi r}_{\min } ^{d+7}}\nonumber\\
&+\frac{(d-2) \left(Q^2 r_{\min } ^8+2    r_{\min } ^{2 d} \left( \left(30-11 d+d^2\right) a    +\left(12-7 d+d^2\right) r_{\min } ^2\right)\right)  \Omega   _{   } \epsilon }{32{  }{\pi r}_{\min } ^{d+7}} \nonumber\\
&+O(\epsilon)^2.  \label{eq3.34}
\end{align}

Substituting Eq.(\ref{eq3.34}) into Eq.(\ref{eq3.31}), we have
\begin{align}
{df}_{\min }&=-\frac{ 2 \left(2-3 d+d^2\right) r_{\min } ^{4+2 d}{dr}_{\min } }{2 \left((d-2) l^2 \left(2 a    +r_{\min } ^2\right)\right)r_{\min } ^{2 d+1}}\nonumber\\
&-\frac{ l^2 \left(-Q^2 r_{\min } ^8+2 (d-2)     r_{\min } ^{2 d} \left(\alpha (d-5)    +(d-3) r_{\min } ^2\right)\right){dr}_{\min } }{2 \left((d-2) l^2 \left(2 \alpha    +r_{\min } ^2\right)\right)r_{\min } ^{2 d+1}}\nonumber\\
&-\frac{ -2 r_{\min } \left( l^3 Q r_{\min } ^8{dQ}-(d-2) r_{\min } ^{2 d} \left((d-5) {d\alpha}    ^2 l^3-2 (d-1) {dl} r_{\min } ^4\right)\right) \epsilon }{2 \left((d-2) l^3 \left(2 a    +r_{\min } ^2\right)\right)r_{\min } ^{ 2d+2}}\nonumber\\
&-\frac{(d-2){  }l{  }\left(2-3 d+d^2\right) r_{\min } ^{4+2 d}{dr}_{\min }\epsilon }{\left((d-2) l^3 \left(2 \alpha    +r_{\min } ^2\right)\right)r_{\min } ^{2 d+1}}\nonumber\\
&-\frac{l^2 \left(Q^2r_{\min } ^8+2     r_{\min } ^{2 d} \left(\alpha \left(30-11 d+d^2\right)    +\left(12-7 d+d^2\right) r_{\min } ^2\right)\right)\epsilon }{2 \left((d-2) l^3 \left(2 \alpha    +r_{\min } ^2\right)\right)r_{\min } ^{2 d+1}}\nonumber\\
&+O(\epsilon)^2.  \label{eq3.35}
\end{align}
For the equation $f(r_h)=0$, we also can expand and solve it, which leads to lastly
\begin{align}
l=\frac{\sqrt{2} \sqrt{2-3 d+d^2} r_{\min } ^{ ( d+2)}}{\sqrt{Q^2 r_{\min } ^8- \alpha    ^2\left(20 r_{\min } ^{2 d}-14{  }d r_{\min } ^{2 d}+2{  }d^2{  }r_{\min } ^{2 d}\right)-     r_{\min } ^{2+2 d}\left(12-10 d +2 d^2 \right)}}.  \label{eq3.36}
\end{align}
In addition, based on Eq.(\ref{dfrp}), we can get
\begin{align}
&dl=\frac{-\sqrt{2} \sqrt{2-3 d+d^2} r_{\min } ^{2+d}\left( Q r_{\min } ^8{dQ}-\left(10-7 d+d^2\right) {d\alpha}    ^2 r_{\min } ^{2 d}\right)}{\left(Q^2 r_{\min } ^8-2 (d-2)     r_{\min } ^{2 d} \left(\alpha (d-5) +(d-3) r_{\min } ^2\right)\right) ^{3/2}}\nonumber\\
&-\frac{\sqrt{2} \sqrt{2-3 d+d^2} \left((d-2) {dr}_{\min } \left(-Q^2 r_{\min } ^8+2     r_{\min } ^{2 d} \left(2 \alpha (d-5)    +(d-3) r_{\min } ^2\right)\right)\right)}{r_{\min } ^{-1-d} \left(Q^2 r_{\min } ^8-2 (d-2)     r_{\min } ^{2 d} \left(a (d-5)    +(d-3) r_{\min } ^2\right)\right) ^{3/2}}.  \label{eq3.37}
\end{align}
Substituting Eq.(\ref{eq3.36}) and Eq.(\ref{eq3.37}) into Eq.(\ref{eq3.35}), we find
\begin{align}
df_{\min }=O(\epsilon )^2.  \label{eq3.38}
\end{align}
In  \cite{Gwak:2015sua}, it was claimed that $df_{\min }$ can be  neglected for it is the high order terms of $\epsilon $. In fact, $\delta$ is a small quantity, we can not neglect the  contribution of $O(\epsilon )^2$ to $f(r_{\min }+{dr}_{\min })$ for both of them are small.
We can find that $\delta$ is also a function of $\epsilon$.  As we expand $f(r_h)$ at $r_{\min }$ to the second order, we find
\be
f(r_h)=f(r_{\min })+f^{\prime}(r_{\min })\epsilon +\frac{1}{2}f^{\prime\prime}(r_{\min })\epsilon^2+O(\epsilon )^3=0.
\ee
For $f^{\prime\prime}(r_{\min })\neq 0$, $O(\epsilon )^3$ thus can be omitted since the dominant term is that of $\epsilon^2$. In this case,   we have
\bea
\delta&=&-\frac{1}{2}f^{\prime\prime}(r_{\min })\epsilon^2\nonumber\\
&=&-\frac{\alpha \left(d^2-9 d+20\right)   l^2+r_{min}^2 \left(\left(d^2-3 d+2\right) r_{min}^2+(d-3)^2   l^2\right)}{l^2 r_{min}^2 \left(2 \alpha  +r_{min}^2\right)}\epsilon^2.
\ena\label{final}
\begin{figure}[h]
\centering
\includegraphics[scale=0.7]{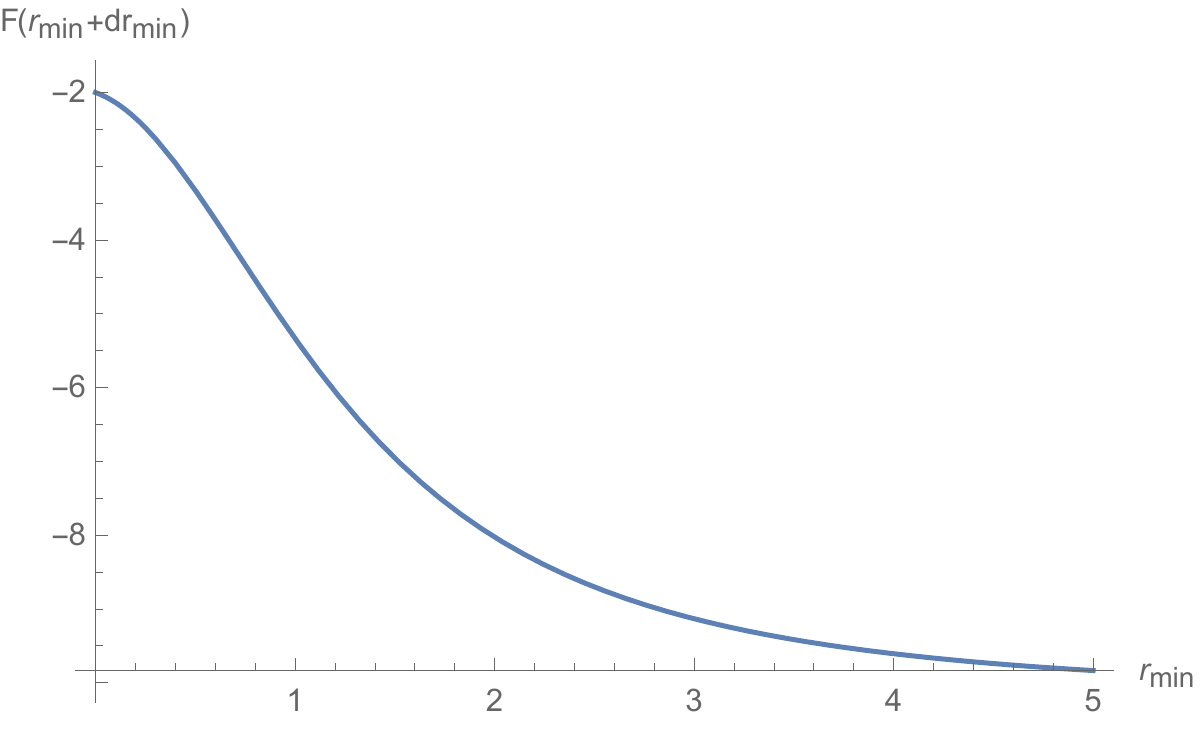}
\caption{The relation between $F(r_{\min }+{dr}_{\min })$ and $r_{min}$ for  $Q=l=\alpha=1, d=5, da=dl=dr=0.08$.}
\label{fig5}
\end{figure}

\begin{figure}[h]
\centering
\includegraphics[scale=0.7]{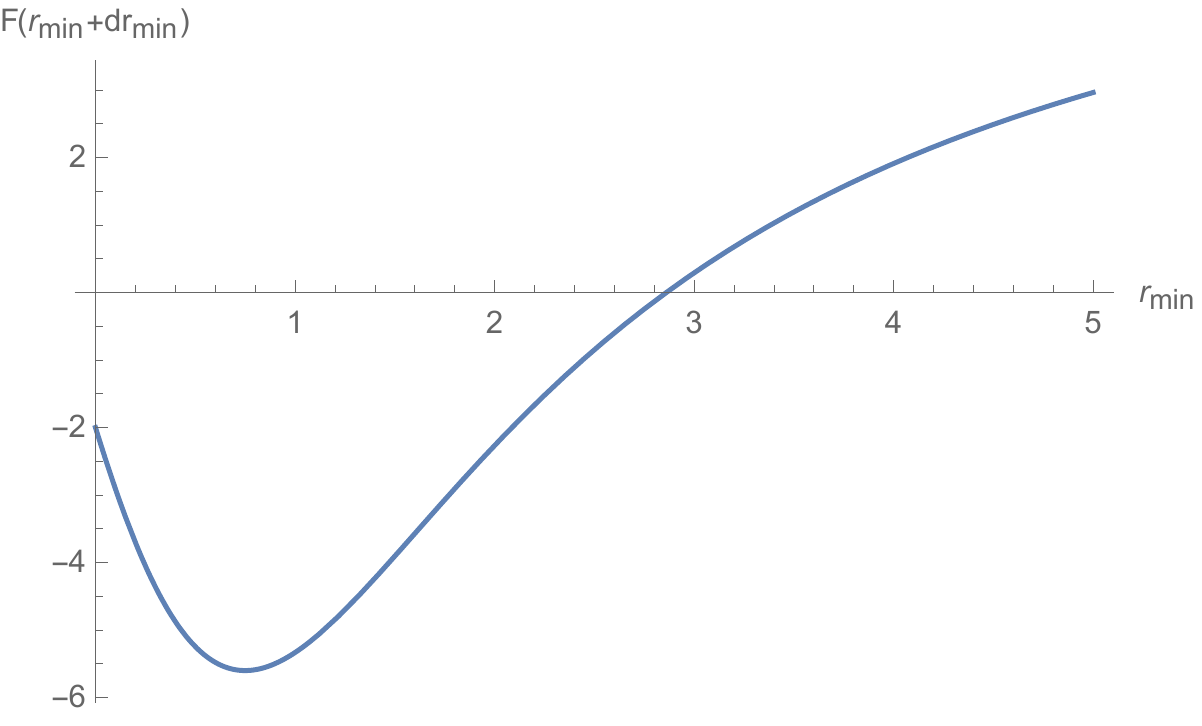}
\caption{The relation between $F(r_{\min }+{dr}_{\min })$ and $r_{min}$ for  $Q=l=\alpha=1, d=5, da=dl=dr=0.8$.}
\label{fig4}
\end{figure}
In addition, to the second order, $df_{\min }$ can be simplified as
\begin{align}
df_{\min }=\frac{X}{l^3 r_{min}^3 \left(2 \alpha+r_{min}^2\right)} \epsilon^2 + O(\epsilon )^3,  \label{dfef}
\end{align}
in which
\bea
X&=&\alpha \left(16 d^2-d^3-83 d+140\right)  {dr_{min}} l^3-\left(d^2-9 d+20\right)  {d\alpha} l^3 r_{min}+2 \left(d^2-3 d+2\right) {dl} r_{min}^5\nonumber\\
&+&(3-d) \left(d^2-8 d+15\right)  {dr_{min}} l^3 r_{min}^2+(3-d) \left(d^2-3 d+2\right) {dr_{min}} l r_{min}^4.
\ena
In this case, we can obtain the value of  the final state of function $f$. For simplicity, we will discuss  $F(r_{\min }+{dr}_{\min })\equiv f(r_{\min }+{dr}_{\min })/\epsilon^2$, that is

\bea
F(r_{\min }+{dr}_{\min })&= &\frac{X}{l^3 r_{min}^3 \left(2 \alpha+r_{min}^2\right)}\nonumber\\
&-&\frac{\alpha \left(d^2-9 d+20\right)   l^2+r_{min}^2 \left(\left(d^2-3 d+2\right) r_{min}^2+(d-3)^2   l^2\right)}{l^2 r_{min}^2 \left(2 \alpha  +r_{min}^2\right)}.
 \ena
 For different values of the parameters $\alpha, l, r_{min}, d, d\alpha, dl, dr_{min}$, the configurations of $F(r_{\min }+{dr}_{\min })$ are different. In this paper, we find $d\alpha, dl, dr_{min}$ affect the configuration drastically. For the case $d\alpha= dl=dr_{min}=0.08$, the configuration of  $F(r_{\min }+{dr}_{\min })$ is plotted in Figure  \ref{fig5}. We can see that $F(r_{\min }+{dr}_{\min })$ is negative, implying that there are horizons always. And for the case $d\alpha= dl=dr_{min}=0.8$, the configuration of  $F(r_{\min }+{dr}_{\min })$ is plotted in Figure  \ref{fig4}.
  We can see that $F(r_{\min }+{dr}_{\min })$  may be positive. In this case, there is not a horizon to cover the singularity and the   weak cosmic censorship conjecture is violated.
As we change the values of the parameters $\alpha, l, r_{min}, d, d\alpha, dl, dr_{min}$,  the configurations of $F(r_{\min }+{dr}_{\min })$ will change too. However, we can find that $F(r_{\min }+{dr}_{\min })$ is positive always for some values of the parameters.
 So, in the extended phase space, the  weak cosmic censorship conjecture is violable, depending on the values of  $\alpha, l, r_{min}$ and their variations.

\section{Discussion and conclusions}\label{sec:5}
As a charged particle drops into the Gauss-Bonnet AdS black hole, we  obtained the energy-momentum relation   near the horizon
via the  Hamilton-Jacobi equation.   We found that there was a relation between the energy, momentum, and chemical potential, which is conjectured to be the first law of black hole thermodynamics. To confirm this conjecture, we investigated the variation of the event horizon with the help of energy conservation as well as  charge conservation and  found that the energy-momentum relation was noting but the first law of thermodynamics  in both the normal phase space and extended phase space.

With the variation of the event horizon, we also checked the second law of thermodynamics by investigating the variation of entropy. In the normal phase space, we found that the second law was valid for the  variation of entropy was positive always. This conclusion is independent of the  Gauss-Bonnet coefficient. In the extended phase space, the variation of the entropy is more sophisticated. We found that there was always a phase transition point, which divides the variation of entropy into positive and negative region. The variation of entropy is negative for the extremal and near-extremal black holes, while  positive  for the far-extremal black holes. In addition, we found that the phase transition point is independent of the  Gauss-Bonnet coefficient though the value of the variation of entropy depends.
Therefore, we concluded that in the extended phase space, the second law was violated for the extremal and near-extremal black holes.

In the normal and extended phase space, we also investigated the weak cosmic censorship conjecture.  We mainly concentrated on studying  how the minimum value of the function that determine the locations of the horizons  move.
 In the normal phase space, we found that the function are stable and  move downward respectively  for the extremal and near-extremal black holes as a charged particle is  absorbed, which implies that there are horizons always so that the   weak cosmic censorship conjecture
is valid. In the extended phase space,   the validity or violation of the weak cosmic censorship conjecture is more subtle.  We found that for the extremal  Gauss-Bonnet AdS black hole,
the   weak cosmic censorship conjecture
is valid always since the final state of the black hole is also an extremal black hole. While for the near-extremal Gauss-Bonnet AdS black hole, the weak cosmic censorship conjecture was found to be  violable, depending on the values of  $\alpha, l, r_{min}$ and their variations. Our result is different from that in  \cite{Gwak:2015sua}, where  the weak cosmic censorship conjecture for the near extremal black hole was found to be valid.  The reason arises from that they neglected the contribution of the second order term of $\epsilon$ to $df_{\min }$. As we shown, the second order term of $\epsilon$  can not be neglected for the initial state is also a function of $\epsilon^2$. Our result is thus more reasonable.

\section*{Acknowledgements}{This work is supported  by the National
Natural Science Foundation of China (Grant No. 11875095), and Basic Research Project of Science and Technology Committee of Chongqing (Grant No. cstc2018jcyjA2480).}

\end{document}